# The Uncertainty Relation for Quantum Propositions

**Paola Zizzi**

Department of Psychology, University of Pavia,
Piazza Botta, 6, 27100 Pavia, Italy
paola.zizzi@unipv.it

## Abstract

Logical propositions with the fuzzy modality "Probably" are shown to obey an uncertainty principle very similar to that of Quantum Optics. In the case of such propositions, the partial truth values are in fact probabilities. The corresponding assertions in the metalanguage, have complex assertion degrees which can be interpreted as probability amplitudes. In the logical case, the uncertainty relation is about the assertion degree, which plays the role of the phase, and the total number of atomic propositions, which plays the role of the number of modes. In analogy with coherent states in quantum physics, we define "quantum coherent propositions" those which minimize the above logical uncertainty relation. Finally, we show that there is only one kind of compound quantum-coherent propositions: the "cat state" propositions.

# 1. Introduction

A quantum logic should be the logic of a truly quantum system. The latter is described in terms of some conjugate variables, which satisfy some uncertainty relation. It follows that a logic is quantum if and only if the propositions of its language, which are in a one-to-one correspondence with the states of the quantum system, obey themselves some (logical) uncertainty principle. In standard quantum logic [1] the requirement of the (physical) uncertainty principle led to a non-distributive lattice of propositions. However, that was just an algebraic expression of such a requirement. Instead, in this paper we will get an explicit realization of a logical uncertainty principle for quantum propositions.

To start, we will consider atomic propositions of the quantum logical language, or quantum object-language (QOL), which can be asserted, in the quantum metalanguage (QML) [2] with an assertion degree. This fact requires that the atomic propositions in the QOL are endowed with a fuzzy modality "Probably" [3] and have fuzzy (partial) truth-values [4]. The latter, moreover, sum up to one. In general, such a set of probabilistic propositions is a subset of a bigger set, including also non-probabilistic ones. We find an uncertainty relation between the (partial) truth-value and the total number of propositions. Also, we define as "quantum coherent propositions" those propositions which minimize the logical uncertainty relation.

In quantum physics, the concept of coherence can have different meanings. In Quantum Mechanics (QM), coherence is a property of pure states, and is related to quantum superposition and entanglement. There are some particular quantum states of the Quantum Harmonic Oscillator (QHO), which are eigenvectors of the annihilation operator, and have minimal uncertainty. They were formally defined by Schrödinger [5], and are the most classical states of the QHO. In Quantum Field Theory (QFT), more precisely in Quantum Electrodynamics, these states were studied by Glauber and named "coherent states" [6]. Glauber coherent states minimize the phase-number uncertainty principle of quantum optics [7].

There is a difference between the linear superposition of pure states and that of Glauber coherent states. While the former gives rise to another pure state, the latter in general does not produce another Glauber coherent state, unless it is a "cat state" [8].

The concept of coherence in logic has been investigated only at the classical level. That was a restriction which, in our opinion, was greatly responsible of the arising of the well known criticisms. A (classical) coherence theory of truth (see for example Ref. [9]) states that the truth of any (true) proposition consists in its coherence with some specified set of propositions. Then, according to the coherence theory, the truth conditions of propositions consist in other propositions. But a definition of logical coherence itself is not given: the argument is circular.

For a coherence theory of truth, the main criticism, due to Russell [10] is that contradictory beliefs can be shown to be true according to coherence theory.

For *coherentism* in general, or the "coherence theory of justification" (which characterizes epistemic justification as a property of a belief only if that belief is a member of a coherent set) the main criticism is the *regress argument*.

Then, we will avoid then the classical concept of logical coherence, and will focus on quantum logical coherence, for which it is possible, as we will see, to give an operational definition.

In our case, coherent quantum logic must have a physical interpretation at two levels. Coherent quantum propositions should be interpreted physically as quantum mechanical pure states, and their assertions should be interpreted as Glauber coherent states. This is due to the fact that QOL is to be interpreted within QM, while QML is to be interpreted within QFT [2].

The paper is organized as follows.

In Sect. 2, we review the basic notions of (classical) metalanguage, the reflection principle and the (classical) definitional equation for the logical connective & in Basic logic [11]. Then, we briefly introduce the quantum assertions with complex assertion degrees of QML [2], and the quantum definitional equation for the logical connective of quantum superposition.

In Sect. 3, we show that the atomic propositions of a QOL, which are endowed with a fuzzy modality "Probably", obey an uncertainty principle between the partial truth-value and the total number of propositions. Moreover, we give the definition of logical quantum coherence for atomic propositions.

In Sect. 4, we look for compound quantum-coherent propositions. To this aim, we need to modify Tarski Convention T [12] as convention TP (true probably). We find that the only possible quantum-coherent compound propositions are semi-classical propositions whose structure reminds the Schrödinger "cat" state.

In Sect. 5, we briefly review the concept of quantum coherence in QM and in QFT. We argue that logical coherence should have both interpretations, respectively at the level of QOL and at the level of QML. In particular, we identify a sequent labelled by an assertion degree with a coherent state.

Sect. 6 is devoted to the conclusions.

## 2. Quantum Metalanguage

A metalanguage (ML) is a language which talks about another language, called object-language (OL).

A (classical) formal ML consists of (classical) assertions, and meta-linguistic links among them. (By classical assertions, we mean assertions which are stated with certitude). It consists of:

i) Atomic assertions: $\vdash A$ (A declared, or asserted), where A is a proposition of the OL.

ii) Meta-linguistic links: $\vdash$ ("yelds", or "entails"), and (metalinguistic "and").

iii) Compound assertions. Example: $\vdash A$ and $\vdash B$.

Let us consider the introduction of the logical connective & in Basic logic [11]

In the OL, let A, B be propositions.

In the ML, I read: A decl. , B decl, that is: $\vdash A$ , $\vdash B$ respectively (where "decl." is the abbreviation of "declared", which also can mean "asserted").

Let us introduce a new proposition A&B in the OL. In the ML, we will read: A&B decl., that is: $\vdash A \& B$.

The question is: From A &B decl., can we understand A decl. and B decl. ?

More formally, from $\vdash A \& B$ can we understand $\vdash A$ and $\vdash B$ ?

To be able to understand A decl. and B decl. from A&B decl, we should solve:

$$\vdash A \& B \quad \underline{\text{iff}} \quad \vdash A \quad \underline{\text{and}} \quad \vdash B \tag{2.1}$$

which is the definitional equation of the connective $\&$ in Basic logic [11].

A quantum metalanguage (QML) [2] consists of:

i) Quantum atomic assertions: $\vdash^{\lambda} p$ \tag{2.2}

where $p$ is a proposition of the quantum object-language (QOL), and $\lambda$ is a complex number, called the assertion degree, which indicates the degree of certitude in stating the assertion. In the limit case $\lambda = 1$, quantum assertions reduces to classical ones. The truth-value of the corresponding proposition $A$ in the QOL, is given by:

$$v(p) = |\lambda|^2 \in [0,1] \tag{2.3}$$

which is a partial truth-value as in Fuzzy Logic [4].

ii) Meta-linguistic links: $\vdash$ ("yelds", or "entails"), and (metalinguistic "and"), as in the classical case.

iii) Compound assertions. Example: $\vdash^{\lambda_0} p_0$ and $\vdash^{\lambda_1} p_1$

iv) Meta-data:

$$\sum_{i=0}^{n-1} v(p_i) = 1 \tag{2.4}$$

where $n$ is the number of atomic propositions in the QOL.

As in the classical case, one should solve the definitional equation of the quantum connective $\&_{\lambda_1}^{\lambda_0}$ [2]:

$$\left|- p_0 \,_{\lambda_0} \&_{\lambda_1} p_1 \quad \underline{\text{iff}} \quad \left|-^{\lambda_0} p_0 \quad \underline{\text{and}} \quad \left|-^{\lambda_1} p_1 \tag{2.5}$$

with the constraint:

$$\left|\lambda_0\right|^2 + \left|\lambda_1\right|^2 = 1 \tag{2.6}$$

which is the meta-data in Eq. (2.4) written in terms of the assertion degrees, by the use of Eq. (2.3).

## 3. The uncertainty principle for quantum propositions

Let us consider a set S of N atomic Boolean propositions:

$$\psi_i \quad _{(i=1,2,\dots\dots N)}$$

Let us call $p_i$ $_{(i=1,2\dots\dots n)}$ with $n < N$, the propositions of a subset $S' \subset S$, to which it is possible to assign a probability $p$ such that $\sum_{i=1}^{n} p(p_i) = 1$.

There is a relation between the probability $p$ and the fuzzy notion *probably*. In fact, the latter can be axiomatized as a *fuzzy modality* [3] as follows. Having a probability $p$ on Boolean formulas, define for each such formula $p_i$ a new formula $P(p_i)$, read "probably $p_i$", and define the truth value of $P(p_i)$ to be the probability of $p_i$ [3], that is: $v(P(p_i)) = p(p_i) \in [0,1]$. Then, it holds:

$$\sum_{i=1}^{n} v(P(p_i)) = 1 \tag{3.1}$$

Let us call $\varphi_i$ $_{(i=n+1,\dots\dots n+r=N)}$ the remaining $r = N - n$ Boolean propositions to which it was not assigned a probability. The $\varphi_i$ belong to the complement of S' in S. Assuming the propositions $\varphi_i$ true, with full truth value 1, it holds:

$$\sum_{i=n+1}^{N=n+r} v(\varphi_i) = r \tag{3.2}$$

From (3.1) and (3.2) it follows:

$$\sum_{i=1}^{N} v(\psi_i) = 1 + r \tag{3.3}$$

In the limit case where all the propositions had the same truth value $v$, Eq. (3.3) would give:

$$v \cdot N = 1 + r \tag{3.4}$$

Moreover, in the particular case with $r = 0, \quad N = n$, Eq. (3.4) gives:

$$v = \frac{1}{n} \tag{3.5}$$

Then, we are allowed to formulate the following *uncertainty relation*:

"It is impossible to fully determine both the truth values of the propositions belonging to a set S, and the power of S":

$$\Delta v \cdot \Delta N \geq 1 \tag{3.6}$$

where $\Delta v$ and $\Delta N$ indicate the uncertainties of the truth value and of the power of the set, respectively.

Notice that for $r = 0, \quad N = n$, Eq. (3.6) is saturated:

$$\Delta v \cdot \Delta n = 1 \tag{3.7}$$

The propositions $\psi_i$ $_{(i=1,2,......N)}$ , with truth-value $v$ in the object-language, are assertions in the metalanguage, with assertion degree $\lambda$, such that $v = |\lambda|^2$. Notice that for Boolean propositions it is: $v = 1 = \lambda$, while for probabilistic propositions we have: $\lambda \in C$ and $v \in [0,1]$.

Then, in general, the uncertainty of the truth value $\Delta v$ can be expressed in terms of the uncertainty of the assertion degree $\Delta \lambda$ as: $\Delta v = 2|\lambda|\Delta \lambda$, and Eqs. (3.6) and (3.7) can be rewritten, respectively, as:

$$\Delta \lambda \cdot \Delta N \geq \frac{k}{2} \qquad (3.8)$$

$$\Delta \lambda \cdot \Delta n = \frac{k}{2} \qquad (3.9)$$

where $k$ is a constant: $k = \frac{1}{|\lambda|}$.

Eq. (3.8) reminds of the uncertainty relation phase-number [7] of quantum optics, which is saturated by coherent Glauber states.

We give the following definition: "Quantum-coherent atomic propositions are those fuzzy-probabilistic atomic propositions for which the partial truth values are all equal and sum up to 1".

Notice that the mean of the partial truth values of the propositions $P(p_i)$:

$$\bar{v} = \frac{\sum_{i=1}^{n} v_i}{n} \qquad (3.10)$$

is, because of the meta-data in Eq. (2.4), equal to the truth value of coherent propositions given in Eq. (3.5):

$$\bar{v} = v_{coh} = \frac{1}{n} \qquad (3.11)$$

## 4. Compound quantum-coherent propositions

Metalinguistic sequents are assertions in the metalanguage. There are close relations between assertions, the truth values of propositions in the object-language, and the truth predicate of Tarski [12 ], the latter being formulated in the metalanguage. However, when the certitude in the assertion is not full, also the truth values of the propositions are partial, and the truth predicate of Tarski must be modified.

By Tarski Convention T, every sentence p of the object-language OL must satisfy:

(T): 'p' is true iff p

where 'p' stands for the name of the proposition p, which is the translation in the metalanguage ML, of the corresponding proposition in the OL, and "iff" stands for "if and only if".

For any sentence "probably p" (P(p)), we can reformulate Tarski Convention T as convention (TP) [13] as follows.

(TP): 'p' is probably true iff P(p).

The expression "is probably true" means that the truth of a proposition is asserted with uncertainty, not with complete certitude. The truth predicate has been modified by probability.

In terms of assertions, convention TP reads:

$$\vert\!-^{\lambda} \; 'p' \quad \underline{\text{iff}} \quad P(p) \qquad (4.1)$$

which means that the proposition $'p'$ is asserted with assertion degree $\lambda$ if and only if "probably $p$", with probability $|\lambda|^2 \in [0,1]$, and the partial truth value of $P(p)$ is just the probability of $p$, that is:

$$v(P(p)) = p(p) = |\lambda|^2 \tag{4.2}$$

Let us consider two propositions $p_0, p_1$ of the QOL.

We get, from Convention TP:

$$\mid\!-^{\lambda_0} \, 'p_0' \quad \underline{\text{iff}} \quad P(p_0)$$

and:

$$\mid\!-^{\lambda_1} \, 'p_1' \quad \underline{\text{iff}} \quad P(p_1)$$

with: $v(P(p_0)) = p(p_0) = |\lambda_0|^2$, $v(P(p_1)) = p(p_1) = |\lambda_1|^2$, together with the constraint:

$$v(P(p_0)) + v(P(p_1)) = 1 \tag{4.3}$$

which is the meta-data in Eq. (2.4).

Let us now form, in the QOL, the conjunction $\&^*$ of the two propositions $p_0, p_1$ such that it takes into account the probabilistic nature of $p_0, p_1$. We define then $p_0 \&^* p_1 \overset{def}{=} P(p_0) \& P(p_1)$.

We put:

$$P(p_0) \& P(p_1) \equiv p_0 {}_{\lambda_0}\&_{\lambda_1} p_1 \tag{4.4}$$

where the suffixes $\lambda_0, \lambda_1$ in ${}_{\lambda_0}\&_{\lambda_1}$ indicates the weights by which the two propositions contribute to the logical conjunction.

The identification $p_0 {}_{\lambda_0}\&_{\lambda_1} p_1 = P(p_0 \& p_1)$ can be made only in the particular case with $\lambda_0 = \lambda_1$, and, by the constraint (5.3) it follows $\lambda_0 = \lambda_1 = \dfrac{1}{\sqrt{2}}$, then we have:

$$P(p_0 \& p_1) \equiv p_0 {}_{\frac{1}{\sqrt{2}}}\&_{\frac{1}{\sqrt{2}}} p_1 \tag{4.5}$$

In this particular case, we can apply Convention PT to the new formed proposition $P(p_0 \& p_1)$:

$$\mid\!-^{\lambda = \frac{1}{\sqrt{2}}} (p_0 \& p_1) \quad \underline{\text{iff}} \quad P(p_0 \& p_1) \tag{4.6}$$

We define a compound quantum coherent proposition as the one whose assertion can be derived through Convention PT from the conjunction of two coherent atomic propositions.

In the general case, with $\lambda_0 \neq \lambda_1$, Convention T gives instead:

$$\mid\!- p_0 {}_{\lambda_0}\&_{\lambda_1} p_1 \quad \underline{\text{iff}} \quad P(p_0) \& P(p_1) \tag{4.7}$$

In the general quantum case, the definitional equation of the quantum connective ${}_{\lambda_0}\&_{\lambda_1}$ of quantum superposition was given in Eq. (2.5).

In the particular case of coherent propositions, the definitional equation of the connective ${}_{\frac{1}{\sqrt{2}}}\&_{\frac{1}{\sqrt{2}}}$ of the "cat state" (which is a semi-classical version of the classical connective $\&$) is:

$$\mid\!-P_0 {}_{\frac{1}{\sqrt{2}}}\&_{\frac{1}{\sqrt{2}}} P_1 \quad \underline{\text{iff}} \quad \mid\!-^{\frac{1}{\sqrt{2}}} p_0 \quad \underline{\text{and}} \quad \mid\!-^{\frac{1}{\sqrt{2}}} p_1 \tag{4.8}$$

## 5. The physical interpretation

The physical interpretation of quantum logical coherence should take into account the concept of quantum coherence in Quantum Mechanics (QM) as well as in Quantum Field Theory (QFT). This is due to the fact that the quantum object-language (QOL) must be interpreted in QM, while the quantum metalanguage (QML) must be interpreted in QFT [14] [15]. More precisely, QML is a non-computational aspect of the mind [13], which is strictly related to brain processes described by a dissipative quantum field theory (DQFT) [16] [17]. Moreover, QML can be viewed as a quantum control on the quantum mind, or on a quantum robot [18] [19] whose logic is the QOL. In this context, it would be interesting to investigate about the particular role of that sector of the QML corresponding to quantum-coherent propositions in the QOL.

Quantum coherence in QM is a property of pure states, whose linear superposition is also a pure state. More specifically: given n pure states $|\psi_i\rangle$ $(i = 1, 2, \ldots, n)$, the superposed state $|\Psi\rangle = \sum_{i=1}^{n} c_i |\psi_i\rangle$ is also a pure state, with:

$$c_i \in C, \qquad \sum_{i=1}^{n} |c_i|^2 = 1 . \tag{5.1}$$

The logical equivalent of Eq. (5.1) is Eq. (3.1) for fuzzy propositions with modality P.

Quantum coherence in QFT is a property of some particular quantum field states, the coherent states $|\alpha\rangle$, which are eigenvectors of the annihilation operator $\hat{a}$, with eigenvalues $\alpha$:

$$\hat{a}|\alpha\rangle = \alpha|\alpha\rangle . \tag{5.2}$$

As the operator $\hat{a}$ is non-hermitian, the eigenvalue $\alpha$ is in general a complex number.

The metalogic equivalent of Eq. (5.2 ) is the sequent: $\vdash^{\alpha} p_\alpha$ , with $\alpha \in C$ .

Coherent states minimize the position-momentum uncertainty relation: $\Delta x \cdot \Delta p = \dfrac{\hbar}{2}$ .

In particular, Glauber coherent states [6] minimize the phase-number [7] uncertainty relation:

$$\Delta \varphi \cdot \Delta n = \frac{1}{2} \tag{5.3}$$

in quantum optics.

Notice the similarity between Eq. (5.3) for coherent optical states and Eq. (3.9) for quantum coherent propositions.

Finally, the superposition of two coherent states $\alpha|\alpha\rangle + \beta|\beta\rangle$ is not, in general a coherent state, unless $\alpha = \beta = \dfrac{1}{\sqrt{2}}$ . The resulting coherent state:

$$\frac{1}{\sqrt{2}}(|\alpha\rangle + |-\alpha\rangle) \tag{5.4}$$

is called the "cat" coherent state [8] in quantum optics.

In our case, in order to give a physical interpretation to quantum coherent propositions, one should consider "generalized" coherent states [20] [21], that is, field states which are eigenvectors of a general non-hermitian operator, which plays the role of the annihilation operator.

We would then restrict ourselves to two non-hermitian operators, which are $2 \times 2$ matrices:

$$A = \begin{pmatrix} \alpha & 0 \\ 0 & 0 \end{pmatrix} \qquad B = \begin{pmatrix} 0 & 0 \\ 0 & \beta \end{pmatrix} \quad \text{with } \alpha, \beta \in C \tag{5.5}$$

where $A$ is the annihilation operator of the state $|1\rangle = \begin{pmatrix} 0 \\ 1 \end{pmatrix}$ , and the state $|0\rangle = \begin{pmatrix} 1 \\ 0 \end{pmatrix}$ is the eigenvector of $A$ with complex eigenvalue $\alpha$ :

$$A|1\rangle = 0$$
$$A|0\rangle = \alpha|0\rangle \tag{5.6}$$

In the same way, $B$ is the annihilation operator of the state $|0\rangle$, and the state $|1\rangle$ is the eigenvector of $B$ with complex eigenvalue $\beta$ :

$$B|0\rangle = 0$$
$$B|1\rangle = \beta|1\rangle \tag{5.7}$$

For $\alpha = \beta = 1$, the operators $A$ and $B$ reduce, respectively, to the two projector operators $P_0 = \begin{pmatrix} 1 & 0 \\ 0 & 0 \end{pmatrix}$ and $P_1 = \begin{pmatrix} 0 & 0 \\ 0 & 1 \end{pmatrix}$ of the two-dimensional complex Hilbert space $C^2$. In standard quantum logic [1], the two projectors are atomic propositions. Then, the expressions $P_0 |0\rangle = |0\rangle$, $P_1 |1\rangle = |1\rangle$ are assertions. The first one asserts that the state $|0\rangle$ has been measured with probability 1, that is, proposition $P_0$ is asserted with certitude (assertion degree $\alpha = 1$). It corresponds to the classical sequent $|-P_0$. The second one asserts that the state $|1\rangle$ has been measured with probability 1, that is, proposition $P_1$ is asserted with certitude (assertion degree $\beta = 1$). It corresponds to the classical sequent $|-P_1$.

Eqs. (5.6) and (5.7) assert with uncertainty propositions $P_0$ and $P_1$, with assertion degrees $\alpha$ and $\beta$ respectively, and thus correspond respectively to the quantum sequents: $|-^{\alpha} P_0$, $|-^{\beta} P_1$.

Now, let us go back to physics, and look for a third coherent state which is a linear superposition of two coherent states. We should consider a field state $|\Psi\rangle$ which is:

i) a linear superposition of the coherent states $|0\rangle$ and $|1\rangle$, that is, $\alpha |0\rangle + \beta |1\rangle$ with: $|\alpha|^2 + |\beta|^2 = 1$.

ii) proportional to the eigenvector $|\gamma\rangle$ of a non-hermitian operator $C$ with a given eigenvalue $\gamma$: $C|\gamma\rangle = \gamma |\gamma\rangle$.

The above two conditions require that $\gamma = \alpha = \beta = \frac{1}{\sqrt{2}}$, so that: $|\Psi\rangle = \frac{1}{\sqrt{2}} (|0\rangle + |1\rangle)$, which is the qubit "cat state". Then, we have:

$$C|\gamma\rangle = \frac{1}{\sqrt{2}} |\gamma\rangle \qquad (5.8)$$

with:

$$C = \begin{pmatrix} \dfrac{1}{\sqrt{2}} & 0 \\ 0 & \dfrac{1}{\sqrt{2}} \end{pmatrix} \text{ and } |\gamma\rangle = \begin{pmatrix} 1 \\ 1 \end{pmatrix}$$

Eq. (5.8) is the algebraic expression of the quantum sequent:

$$|-^{\frac{1}{\sqrt{2}}} (P_0 \, \& \, P_1) \qquad (5.9)$$

or, equivalently:

$$|-(P_0 \, _{\frac{1}{\sqrt{2}}} \& \, _{\frac{1}{\sqrt{2}}} P_1) \qquad (5.10)$$

Eq. (5.9) (or equivalently Eq. (5.10)) assert the quantum-coherent logical qubit cat state.

We remind that the condition for having a quantum coherent compound proposition was $P(P_0) \& P(P_1) = P(P_0 \& P_1)$ which holds if and only if the two atomic propositions are asserted with the same assertion degree. This is algebraically expressed by: $\alpha P_0 + \beta P_1 = \gamma I_2$, where $I_2$ is the identity matrix of $C^2$.

## 6. Conclusions

In this paper, we have formulated an uncertainty principle for (quantum) logical propositions. To us, this is the criterion for stating that a logic is truly quantum. In fact, standard quantum logic was not able to achieve such a requirement, and then it cannot be considered a proper quantum logic in the above sense.

The problem of standard quantum logic stands in the fact that it just consists of syntax, and is missing the semantics, where the quantum meaning arises. Semantics (or metalanguage) is in fact the seed of any physical interpretation of logic. It is the (quantum) metalanguage that states the quantum logical features of the object-language. More precisely, the assertion degrees of the assertions of the QML are the symptoms that the propositions of the QOL are probabilistic and fuzzy at the same time. This leads to the uncertainty principle between the fuzzy truth-value and the power of the set of atomic propositions. Related to any uncertainty principle, there is always the Glauber concept of coherence, that is, the saturation of the uncertainty principle itself. Quantum propositions which minimize the above logical uncertainty principle are then defined "quantum-coherent" propositions. The latter are then the most classical propositions in the QOL.

Moreover, coherent states are the most robust states with respect to damping, as they are "pointer states". Any other quantum states, for example Fock states, are more fragile in decoherence, and will decay faster. Thus, quantum atomic assertions in the QML, corresponding to coherent states in the dissipative QFT of the brain, are very robust against a possible reduction to classical logic. Also, their semi-classical nature suggests that they are the best suited for controlling the object-language at the border between the quantum level and the classical one.

Then, we showed that the compound propositions which are quantum-coherent are only the "cat state" propositions, which can be true or false with the same probability. Logical quantum coherence is then the antithesis of (classical) logical consistency.

However, coherent "cat" states, differently from single coherent states, are quite fragile against dissipation. Decoherence transforms the initial pure state (5.4) into a statistical mixture of the two coherent states $|\alpha\rangle$ and $|-\alpha\rangle$. This implies that cat state assertions in the QML, which provide the semi-classical connective $\frac{1}{\sqrt{2}} \& \frac{1}{\sqrt{2}}$ in the QOL through the reflection principle, are created and destroyed very quickly in the brain. The reflection principle, more precisely, the definitional equation of the quantum conjunction, can be viewed as the reduction of the dissipative QFT of the brain to the QM of the quantum-computing mind [15]. The above considerations suggest that we can make only extremely short journeys in our own quantum mind, but that is sufficient to create our logical language.

## Acknowledgements

I wish to thank Giuseppe Vitiello and Eliano Pessa for useful discussions.

## References

[1] G. Birkhoff and J. von Neumann, *The Logic of Quantum Mechanics*, Annals of Mathematics, Vol. 37, 823–843 (1936).

[2] P. Zizzi, "From Quantum Metalanguage to the Logic of Qubits". PhD Thesis, arXiv:1003.5976 (2010).

[3] H´ajek P., Godo L., Esteva F., Probability and Fuzzy Logic. In Proc. Of Uncertainty in Artificial Intelligence UAI'95, (Besnard and Hanks, Eds.) Morgan Kaufmann. San Francisco, 237–244 (1995).

[4] L. A. Zadeh et al. *Fuzzy Sets, Fuzzy Logic, Fuzzy Systems*, World Scientific Press (1996).

[5] E. Schrödinger, "Der stetige Übergang von der Mikro- zur Makromechanik", *Naturwissenschaften* **14,** 664-666 (1926).

[6] R. J. Glauber, "Coherent and incoherent states of radiation field", *Phys. Rev.* **131,** 2766-2788 (1963).

[7] J-P. Gazeau, *Coherent States in Quantum Physics*, Wiley-VCH, Berlin (2009).

[8] S. Haroche, J. M. Raimond. *Exploring the quantum : atoms, cavities and photons.* Oxford University Press (2006).

[9] A. R. White, "Coherence Theory of Truth", *Encyclopedia of Philosophy*, Vol.2. Macmillan:130 (1969).

[10] A. C. Benjamin, "Coherence Theory of Truth", p. 58 in Dagobert D. Runes (ed.), *Dictionary of Philosophy*, Littlefield, Adams, and Company, Totowa, NJ (1962).

[11] G. Sambin , G. Battilotti, C. Faggian, "Basic logic: reflection, symmetry, visibility". *The Journal of Symbolic Logic*, **65**, 979-1013 (2000).

[12] A. Tarski, "The semantic conception of truth". *Philosophy and Phenomenological Research*, **4**, 13-47 (1944).

[13] P. Zizzi, "The non-algorithmic side of the mind". In *Workbook of the $4^{th}$ Quantumbionet Workshop, the dawn of quantum biology, Crema, $30^{th}$ September 2011*, p.9.

[14] P. Zizzi, "Quantum Mind from a Classical Field Theory of the Brain". *Journal of Cosmology*, Vol 14 (2011).

[15] P. Zizzi, "When Humans Do Compute Quantum", in: A Computable Universe, Hector Zenil (Ed), Word Scientific Publishing. Forthcoming.

[16] G. Vitiello, "Dissipation and memory capacity in the quantum brain model". *International Journal of Modern Physics B*. 9, 973-989 (1995).

[17] G. Vitiello. *My double unveiled*. Amsterdam: Benjamins (2001).

[18] P. Zizzi, "Quantum Robot: Quantum Mind Control on a Quantum Computer". In A. Sakaji, I. Licata, J. Singh, S. Felloni (Eds.). *New Trends in Quantum Information* 299-318. Roma: Aracne (2010).

[19] E. Pessa, P. Zizzi, "Brain-Computer Interfaces and Quantum Robots". ArXiv. 0909.1508. Contributed paper to the Conference ".mERGERs. Physical and Cognitive Mutations in Humans and Machines", Laval (France), 24-25 April 2009.

[20] A. Perelomov, *Generalized Coherent States and their applications*. Springer, New York (1986).

[21] W. Zhang, D. H.Feng, R. Gilmore , "Coherent states: Theory and some applications". *Rev. Mod. Phys*. **62**, 867-927 (1990).